# Configuration Database for BaBar On-line

R. Bartoldus, A. Salnikov
*Stanford Linear Accelerator Center, Stanford University, Stanford, California 94309*

G. Dubois-Felsmann
*Caltech, 1200 E. California Bl., Pasadena, CA 91125, USA*

Y. Kolomensky
*LBNL, 1 Cyclotron Rd., Berkeley, CA 94720, USA*

(On behalf of the BaBar Computing Group)

The configuration database is one of the vital systems in the BaBar on-line system. It provides services for the different parts of the data acquisition system and control system, which require run-time parameters. The original design and implementation of the configuration database played a significant role in the successful BaBar operations since the beginning of experiment. Recent additions to the design of the configuration database provide better means for the management of data and add new tools to simplify main configuration tasks. We describe the design of the configuration database, its implementation with the Objectivity/DB object-oriented database, and our experience collected during the years of operation.

## 1. INTRODUCTION

The BaBar on-line system uses a number of the databases to keep various information relevant to the data taking [1]. These databases include:
- Conditions database [2], which contains time-dependent data, such as calibrations, geometry, etc.
- Ambient database, which keeps a track of the history of data-taking conditions. This is a simplified conditions database and it is a part of the detector control system [3].
- Configuration database, which keeps settings for the parts of the data acquisition system (DAQ).
- Prompt Reconstruction databases, which provide support for multi-node calibration.

This paper describes the design and implementation of the configuration database. The main purpose of the configuration database is to provide all participants of the DAQ system with the data needed to configure them prior to data taking.

## 2. CONFIGURATION DATABASE DESIGN

### 2.1. Requirements

The main requirements for the configuration database are the following:
1. Provide a support for configuration of the on-line software and hardware when data taking starts.
2. Be able to reconstruct the exact configuration used for in any run taken in the past.
3. Support both standard data taking with the full BaBar on-line system and standalone subsystems running on their test-stands.

Additionally there are certain requirements, which influence both design and implementation, arising from the fact that the configuration database is a vital part of the real-time DAQ system.

### 2.2. Configuration data

Configuration database stores and serves configuration data. Usually configuration data represent detector and software settings for the data taking, such as voltages, thresholds, trigger cuts, etc. The scope of these settings is from the beginning to the end of the data taking, i.e. single run. Different types of runs may require different settings, e.g. physics data taking, cosmics, and calibrations will need different triggers, high voltages, etc. To serve different types of runs there should be a number of "active" sets of configuration data, exactly which set of configuration data is used for the next run is determined by the run type.

### 2.3. Configuration objects

Configuration data in the database are stored in the configuration objects, which are the "atoms" of the configuration database – they are the basic units of management. Single configuration object keeps related data, usually needed to configure a particular piece of hardware or software.

Every configuration object in the database has an identity consisting of the three separate pieces:
1. Class name – non-empty string representing the type of the object.
2. Optional secondary key – string used to distinguish objects of the same type but used for different purposes.
3. Configuration key – a number. Enumerates objects of the same type/secondary key, similar to version.

Object identity can be represented in a simple textual format like "ClassName:SecondaryKey[ConfigKey]", or "ClassName[ConfigKey]" when a secondary key is missing. This representation is used in the figures below. The configuration database provides direct access to all configuration objects with their identity.

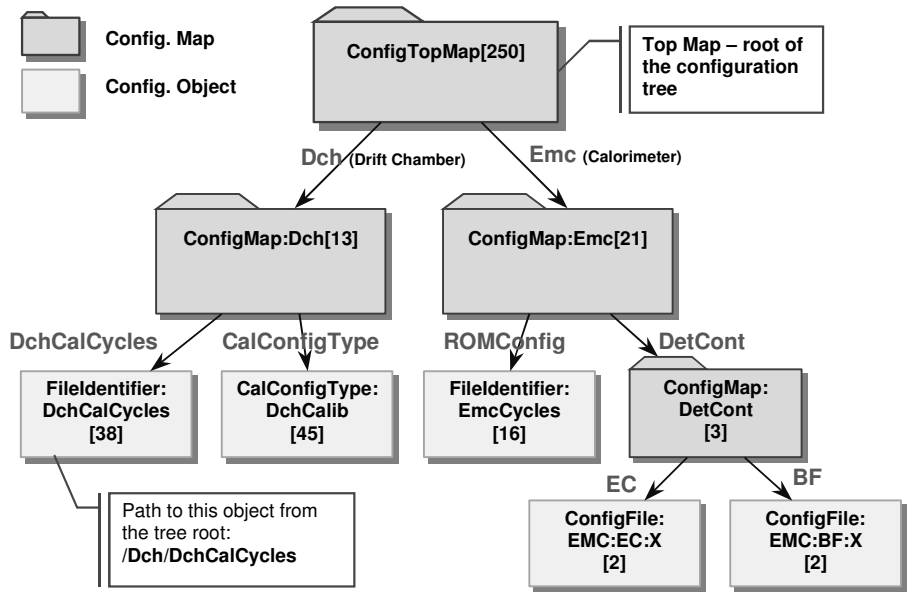

Figure 1: Example of a simple configuration tree. The tree has a root map called "Top Map", which refers to the subsystem maps. Subsystem maps can link to the objects directly or to the next level maps.

In addition to identity, the objects contain actual configuration data. Configuration objects are immutable, once created they never change; this allows exact reconstruction of the past configurations. When the user wants to change something in the configuration data, (s)he needs to create a new configuration object with the same type/secondary key, but with the new configuration key (version). Configuration keys are assigned to the objects by the database itself by incrementing the last used key.

## 2.4. Configuration maps and trees

Complete configuration of the whole system is a potentially big set of all configuration objects needed to setup DAQ system. To simplify management of the configuration this set should be further organized into a single entity.

Configuration maps are special configuration objects, which serve as containers with the named links to other maps or objects. They have all the properties of the configuration objects, such as identity and immutability. The map object names the objects it is referring to; the scope of these names is the map objects itself. This allows two different map objects to refer to the same object with different link names.

The maps are used as the building blocks for the configuration trees. The tree has a single root map, and the identity of the tree is the same as the identity of its root map object. Any configuration objects in the tree can be reached from the root of the tree by its path name, which is the sequence of the names needed to navigate from the root to the object in the tree. Because separate maps can link to the same objects, the trees also can share either basic objects or even sub-trees, thus eliminating the need for duplicating configuration objects. The complete configuration of the whole system is a single tree, and it has an identity, which is the same as the identity of the root map of the tree.

Figure 1 shows an example of very simple configuration tree, which includes only a small number of configuration objects.

There is also a special configuration map in the system which data taking run type strings into the configuration trees. It has links to the root maps, the links are named after the run types. Only one such mapping can be active at any given type, the active map is one with the highest configuration key. The trees referenced from the active run type map are themselves active, and only they can be used in the system. All other trees become a part of the configuration database history.

## 2.5. Accessing configuration objects

The following scenario describes client access to the configuration data.

The run starting sequence is managed by the Run Control code. At the beginning of the new run Run Control determines the run type, which is usually specified by the operator. Then Run Control accesses the database and uses the special run type map to determine the identity of the configuration tree corresponding to the given run type. Run Control then distributes this identity to all participating DAQ processes. Each process uses this identity to access the corresponding configuration tree. Also each process knows the path name of its configuration object in a tree and uses this path to locate and extract corresponding data object.

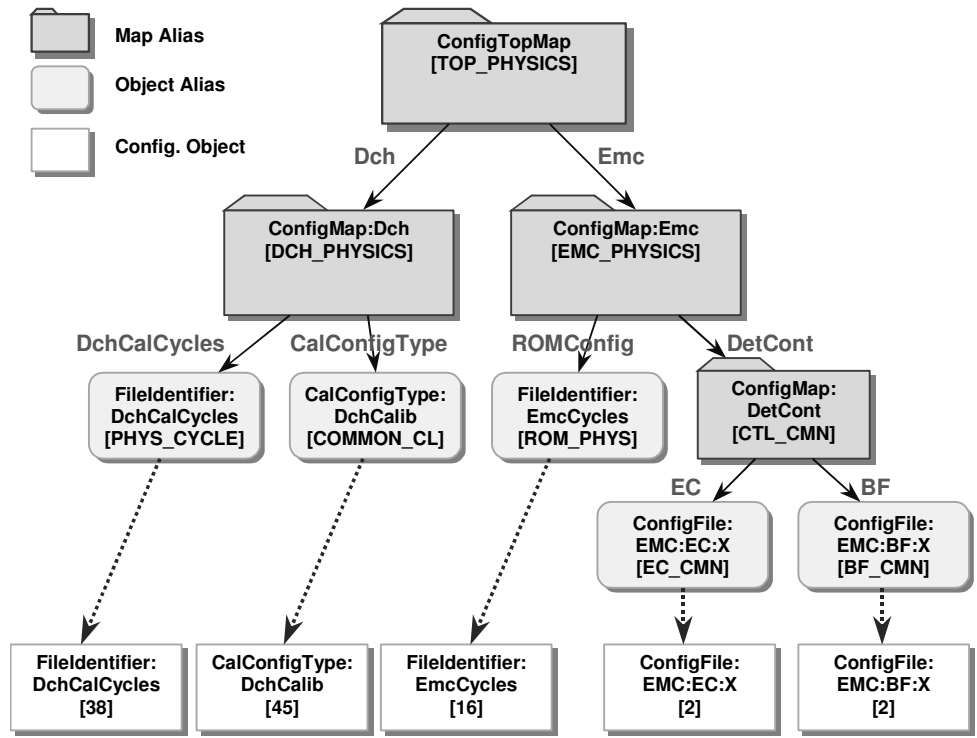

Figure 2: Example of the alias tree. Structure of the aliases repeats the structure of the configuration tree shown on Figure 1. Object aliases have links to the real configuration objects.

## 2.6. Building configuration trees

Configuration trees could be arbitrarily complex, involving large numbers of objects. Building large trees is complicated by two design decisions: 1) it is not possible to change any existing object, 2) many active trees could reference the same basic object or sub-tree.

Typical configuration editing operations in these conditions would become rather involved. For example, replacing one of the leaf objects with a new version requires the update of all maps directly or indirectly connected to this objects in all active trees, and then replacing any modified tree in the run type map with the new tree.

To facilitate the configuration editing operations, one more structure is introduced into the configuration database – alias trees. Alias trees repeat the structure of the configuration trees but instead of real configuration objects they are built with the map aliases and object aliases. These aliases are free of one restriction of the configuration objects – they are allowed to change. Also instead of numeric configuration keys aliases are identified by some meaningful alias names. An example of the alias tree is shown in figure 2. The objects aliases have links to the real configuration objects, while map aliases are only placeholders and have no connection to the corresponding configuration maps. Modifications to the alias trees are much easier, for example the version change of the basic object involves only the link between object alias and that configuration object.

A special procedure is used to update the numeric trees after the changes are applied to alias trees. This procedure makes a node-by-node comparison of the active numeric trees and alias trees, and rebuilds the parts of the numeric trees affected by the changes in alias trees.

## 3. IMPLEMENTATION

The configuration database is a vital part of the DAQ system, and it works in a real-time environment, which imposes strict requirements on performance and the quality of the implementation. The total number of clients accessing configuration data may reach 100, and many of them access data simultaneously during the "configure" transition when a new run starts.

### 3.1. Storage technology

BaBar has chosen Objectivity/DB object database [4] as a storage technology for many of its databases, including the configuration database. There are many benefits, which Objectivity offers:

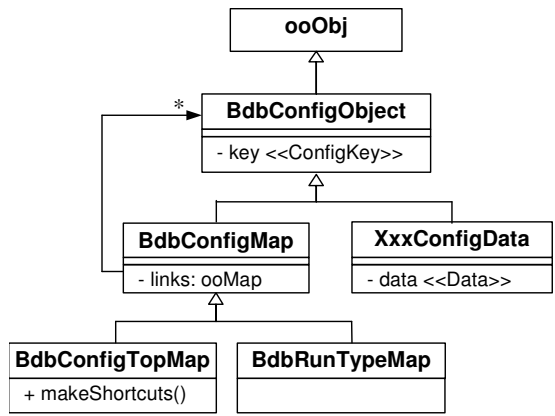

Figure 3: Simplified diagram of the data storage classes. Only one concrete data storage class (XxxConfigData) is shown on this diagram.

- Direct mapping of the persistency constructs into the OO paradigm and C++ constructs (classes).
- Support for inter-object associations with direct links between objects, there is no need to run SQL-like queries to access the objects.
- Complete support of ACID (Atomicity, Consistency, Isolation, and Durability) properties, providing data integrity guarantees.

One more benefit of object databases in the context of the configuration database is the data model. The data model used in object databases is a network of objects, which makes the configuration database design easy to implement in the object database.

Certainly the same design could be implemented with a different data storage technology, such as relational databases, but it would require more effort for such an implementation with probable negative impact on performance.

### 3.2. Data storage classes

All data storage objects in the configuration database are implemented as a single hierarchy of the persistent classes (see Figure 3). The root of the hierarchy is the BdbConfigObject class, which inherits directly from the Objectivity/DB ooObj class. The BdbConfigObject class implements functionality common to all inheriting classes, such as management of configuration keys, storage of bookkeeping information, etc. Some information, such as object class name and secondary key, are the properties of the Objectivity container object, which keeps the objects of the same class.

All concrete classes, which keep configuration data, are subclasses of the BdbConfigObject class, and add specific services for data management.

The configuration map class BdbConfigMap is implemented also as a subclass of BdbConfigObject. The configuration map class has a persistent map instance, which implements the list of named links to other configuration objects. There are two additional subclasses of the BdbConfigMap class, BdbConfigTopMap and BdbRunTypeMap, which implement additional functionality or constrains needed by the top configuration maps and run type maps.

### 3.3. Client data access

Clients of the configuration database use standard BaBar approaches for accessing persistent data [5]. The main idea in these approaches is persistent/transient separation. Client code never manipulates persistent data directly, instead client code is presented with the transient interfaces which provide access to persistent data. This separation allows clients to work independently of the concrete implementation of persistency mechanism. In this model every persistent data class has its transient counterpart. Conversion between transient and persistent representations of configuration data is performed by special proxy objects. The following scenario is used to access the data from client code:

1. Initialization code creates proxy objects for all types of configuration classes used in the client application.
2. When client needs a particular configuration object, it sends a request to the proxy dictionary. Object type (transient object class) is one of the request parameters.
3. Proxy dictionary locates a proxy object responsible for the data of given type and redirects the request to it.
4. Proxy finds persistent data in the configuration database using the parameters of the request or supplied during proxy construction. It then converts the persistent data into transient form and returns the transient object to client.

Both BaBar Framework applications and standalone applications can use proxy mechanism to work with the configuration data.

There are also clients, which have no direct access to database services, such as the code running in VME under control of VxWorks. One additional service was implemented which provides these clients with the configuration data using the BaBar DAQ-specific transport.

### 3.4. Tools and utilities

There are two main tasks in managing the configuration database: 1) creating new configuration objects, 2) modifying configuration trees.

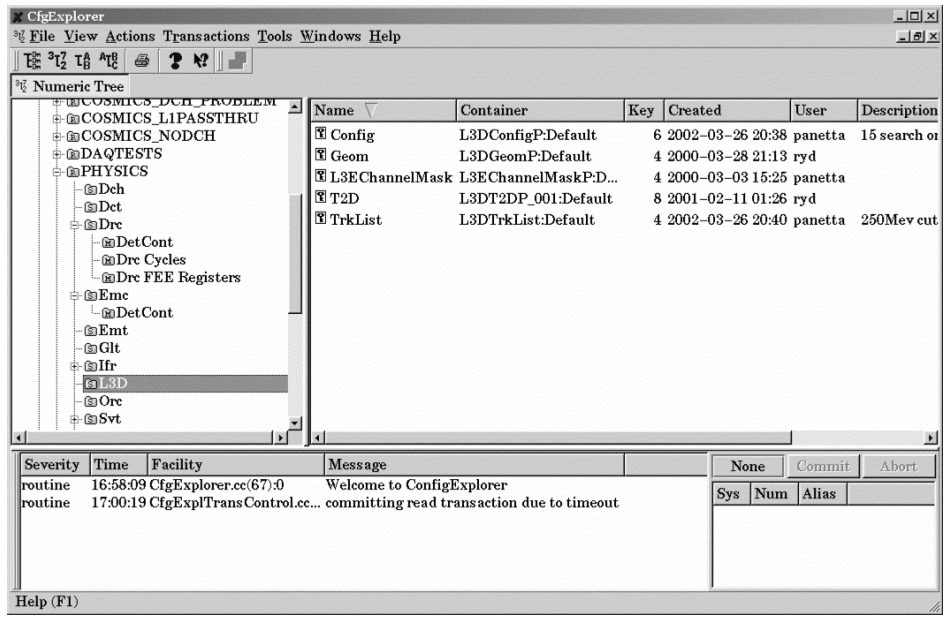

Figure 4: Graphical interface for the configuration database. Upper left panel shows configuration tree structure, right panel displays configuration objects. Bottom part of the window is occupied by the message panel and transaction control.

New configuration objects are created with standalone utilities. Every separate type of configuration objects are created with dedicated utilities, the data for the created object are usually loaded from external sources, such as files, or can be specified as command-line options.

There are two related utilities, which control every aspect of the configuration trees and alias trees. The first is a command-line tool with a simple, but powerful, command language, which makes it easy to write sophisticated command scripts and automation tools. Another one is a GUI application (see Figure 4) built on top of Qt/X11 framework [6]. This application supports the same functionality as the command-line tool.

## 4. CONCLUSION

BaBar has designed and implemented a configuration database for its on-line system, which provides configuration services for the components of DAQ system. The configuration database has successfully operated since the beginning of data taking in 1999, with minor modifications and additions later. The current implementation is based on Objectivity/DB ODBMS. The configuration database is a vital part of the BaBar DAQ system and proved to be sufficiently performant and reliable.

## 5. AKNOWLEDGMENTS

This work is supported by Department of Energy contract DE-AC03-76SF00515.